\documentstyle[epsfig]{elsart}
\begin{document}
\begin{frontmatter}
\title{Testrun results from prototype fiber detectors\\
for high rate particle tracking}

\author{E.C. Aschenauer},
\author{J. B\"ahr},
\author{V. Gapienko\thanksref{russia}},
\author{B. Hoffmann\thanksref{esser}},
\author{H. L\"udecke},
\author{A. Menchikov\thanksref{dubna}},
\author{C. Mertens\thanksref{student}},
\author{R. Nahnhauer}
{\renewcommand{\thefootnote}{\fnsymbol{footnote}}
\hspace{-0.3cm}\footnote{\small corresponding author, phone: +49 33762
    77346, Fax: +49 33762 77330,\\ \hspace*{0.3cm} e-mail: nahnhaue@ifh.de}}
\author{R. Shanidze\thanksref{tbilisi}}
\address{DESY Zeuthen, 15738 Zeuthen, Germany}
{\setcounter{footnote}{0}
\thanks[russia]{on leave from IHEP Protvino, Russia}
\thanks[esser]{now at Esser Networks GmbH, Berlin}
\thanks[dubna]{on leave from JINR Dubna}
\thanks[student]{Summerstudent from University of Clausthal Zellerfeld}
\thanks[tbilisi]{on leave from High Energy Physics Institute, Tbilisi State 
University}}
\begin{abstract}
A fiber detector concept has been realized allowing to registrate
particles within less than 100 nsec with a space point precision of
about 100 $\mu$m at low occupancy. Three full size prototypes have been
build by different producers and tested at a \linebreak 3 GeV electron beam at
DESY. After 3 m of light guides 8 - 10 photoelectrons were
registrated by multichannel photomultipliers providing an efficiency
of more than 99 \%. Using all available data  a resolution of 86 $\mu$m was
measured. 
\end{abstract}
\end{frontmatter}

\section{Introduction}
The advantageous use of fiber detectors for particle tracking has
been demonstrated for very different conditions e.g. in the
UA2-Experiment \cite{lit1}, CHORUS \cite{lit2}, for D0 \cite{lit3} 
and the H1-Forward Proton Spectrometer \cite{lit4}. Due to the
different experimental situations in these applications 
three types of optoelectronic read out techniques are
applied -- Image Intensifiers plus CCD's,
Visible Light Photon Counters (VLPC) and Position Sensitive
Photomultipliers (PSPM). However, all have in common
that the precision of space point measurements is given by hits of
overlapping fibers of several staggered fiber layers. For high rate
experiments demanding online tracking of several hundred particles per
100 nsec bunch crossing such a concept may not work due to too high
occupancy of single fiber channels.

We propose in the following to use overlapping fiber roads composed of
several thin scintillating fibers read out with one clear optical
fiber. The demands and the solutions presented below match to a
possible application of the detector as the inner tracker in the
HERA-B project at DESY \cite{lit5}. Similar ideas have been used by
others \cite{lit6} to build a fiber detector for the DIRAC experiment
at CERN.

\section{Detector Principle}
The fiber detector under discussion is aimed to detect throughgoing
particles with more than 90 \% efficiency within less than 100 nsec
and a
precision of better than 100 $\mu$m. The fibers should not change
their
characteristics significantly after an irradiation of 1 -- 2 Mrad. The
sensitive detector part should have a size of about 25 x 25 cm$^2$. The
scintillating fibers should be coupled to clear optical fibers of
about 3 m length guiding the light to photosensors placed outside the
experimental area.

\begin{figure*}[b]
\vspace*{-0.5cm}
\begin{center}
\epsfig{file=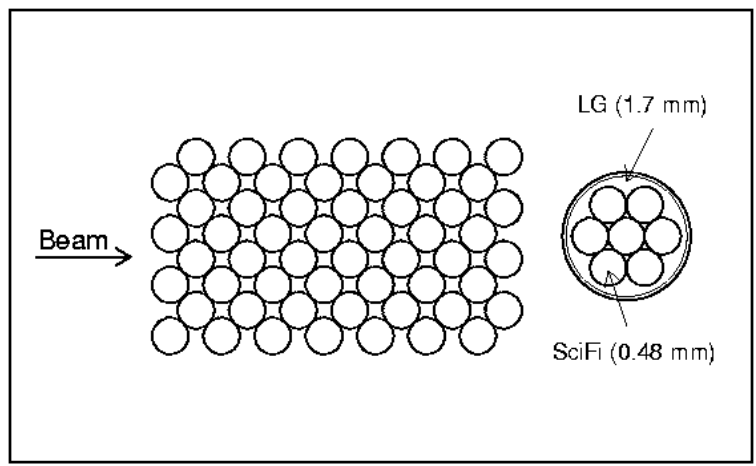,width=10cm}
\vspace*{-2cm}
\caption{\label{bilda} Schematic view of the proposed fiber detector
cross section and coupling principle (LG: light guide fiber)}
\end{center}
\end{figure*}

It is assumed that most particles of interest hit the detector
perpendicular, i.e. with angles less than five degrees with respect to
the beam axis. In this case low occupancy and high light yield  are
guaranteed by using overlapping fiber roads like schematically drawn
in fig. \ref{bilda}. One fiber road consists of several thin
scintillating fibers arranged precisely behind each other and coupled
to one thick light
guide fiber. The scintillating fiber diameter determines the space
point resolution of the detector. The number of fibers per road is
fixed by the scattering angle of particles and the allowed
amount of multiple scattering. It will also influence the factor of
background suppression for tracks with larger inclination or
curvature. The pitch between fiber roads is defined by demanding a
homogeneous amount of fiber material across the detector width.

Keeping in mind the conditions at HERA-B, the following
choices are made:

\begin{center}
\begin{tabular}{l@{$\;=\;$}ll@{$\;=\;$}l}
$\Phi_{fib}$ & 480$\mu$m   & N$_{fib/road}$ &  7 \\
L$_{fib}$    &  30 cm    & p$_{road}$     & 340 $\mu$m\\
$\Phi_{lg}$ & 1.7 mm     & N$_{road}$   &   640 \\  
L$_{lg}$    & 300 cm    & W$_{det}$    &  217.6 mm 
\end{tabular}
\end{center}

with $\Phi$ and L: diameter and length of scintillating and light
guide
fibers, N$_{fib/road}$: number of fibers per road, p$_{road}$:
distance between
neighboured road centers, N$_{road}$: number of roads per detector,
W$_{det}$:
detector width.

The light guide fibers are read out with the new
Hamamatsu\footnote{Hamamatsu Photonics K.K., Electron tube division,
    314--5, Shimokanzo, Tokyooka Village. Iwatagun, Shizuoka--ken,
    Japan} 64
channel PSPM R5900--M64 with a pixel size of 2 x 2 mm$^2$
\cite{lit7}. 
To diminish optical cross talk the thickness of the entrance 
window of the device was decreased to 0.8 mm.

The coupling between scintillating and light guide fibers is done by
loose plastic connectors. The light guides are coupled to the PSPM
using a plastic mask fitting the corresponding pixel pattern.

\section{Material Studies}
Double clad fibers  of three different producers\footnote{BICRON,
  12345 Kinsman Road, Newbury, Ohio, USA} \footnote{pol. hi. tech.,
  s.r.l., S.P. Turanense, 67061 Carsoli(AQ), Italy} \footnote{KURARAY
  Co. LTD., Nikonbashi, Chuo-ku, Tokyo 103, Japan} were tested
concerning light output, light attenuation and radiation hardness for
several fiber diameters and wavelengths of the emitted light. Details
of
these measurements are given in \cite{lit8}. A few results are
summarized below.

\begin{figure*}[t]
\begin{center}
\epsfig{file=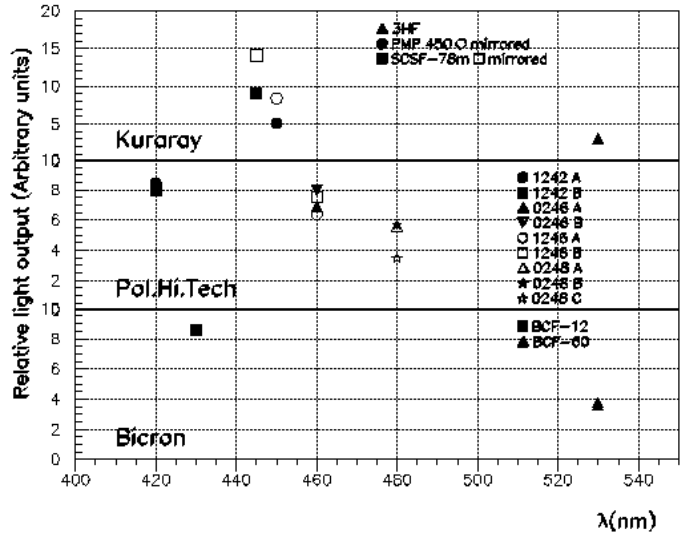,width=10cm}
\vspace*{-0.5cm}
\caption{\label{bildb} Light output from 500 $\mu$m diameter fibers for
            several fiber materials of three producers}
\end{center}
\end{figure*}

The light output of fibers of 500 $\mu$m diameter is shown
in fig. \ref{bildb}. Generally it can be seen, that the light yield
decreases
with increasing scintillator emission wavelength because the PM
sensitivity
curve is not unfolded. There is no remarkable difference between the
best materials of the three producers. A mirror at the end of the
fiber increases the light output by a factor 1.7.

Several tests were performed to couple scintillating and light guide
fibers. Finally the coupling efficiency became better than 95 \%,
independent of the medium between both parts (air, glue, grease).

\begin{figure*}[t]
\begin{center}
\epsfig{file=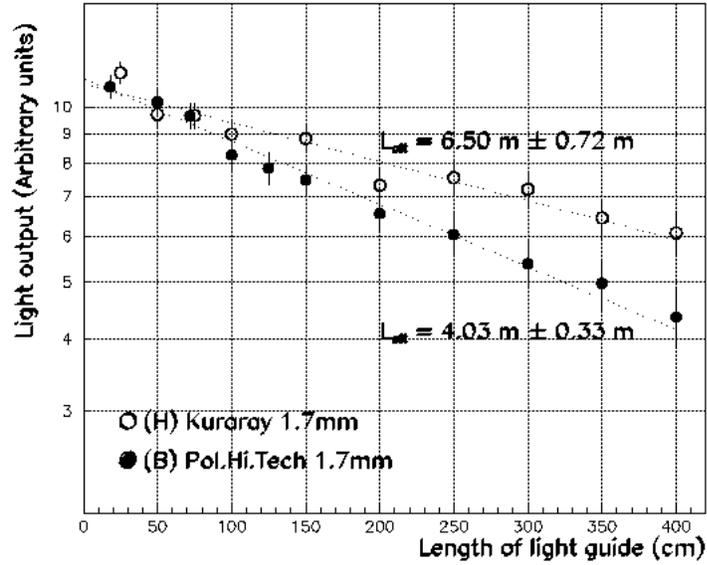,width=10cm}
\vspace*{-1cm}
\caption{\label{bildc} Light attenuation in clear fibers of 1.7 mm diameter
 produced by Kuraray and pol.hi.tech.}
\end{center}
\end{figure*}

The light attenuation of clear fibers was measured coupling them to
single scintillating fiber roads excited by a Ruthenium source. The
clear fibers were cutted back  to the length under
investigation. Results for two producers are given in
fig. \ref{bildc}.

\begin{figure*}
\vspace*{-3cm}
\begin{center}
\epsfig{file=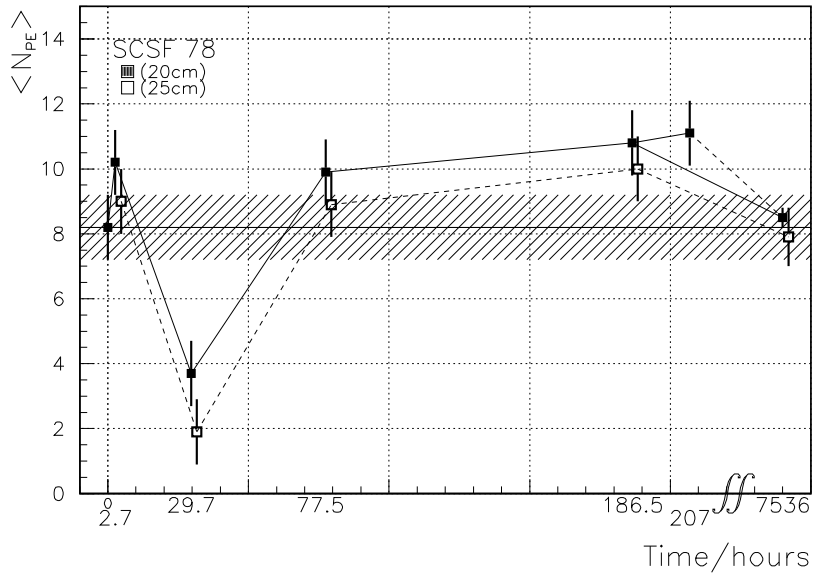,width=12cm}
\vspace*{-5cm}
\caption{\label{bildd} Time evolution of light output for KURARAY SCSF 78M
            fibers irradiated with 0.2 and 1.0 Mrad at 10 and 20 cm
            respectively. The solid and dotted curves
            correspond to measurements with a source placed
            at 20 and 25 cm.}
\end{center}
\end{figure*}

Radiation hardness tests of fibers were made using an intense 70 MeV
proton beam at the Hahn--Meitner--Institute Berlin. 1 Mrad radiation
was
deposited within a few minutes. For all materials investigated we
observed a damage of the scintillator and the transparency of the
fiber which was followed by a long time recovery of up to 600 h. An
example is shown in fig. \ref{bildd}. More detailed studies using glued
and nonglued fibers and irradiate them in air and nitrogen atmosphere are
still ongoing.

Summarizing all results of our material studies we decided to use
the KURARAY fibers SCSF-78M with a diameter of 480 $\mu$m for the
scintillating part of our detector prototypes. For clear fibers still
two choices seem to be possible: 1.7 mm fibers from KURARAY or
pol. hi. tech.. 

Similar investigations of the same materials have been done by our
colleagues from Heidelberg University \cite{lit9}. For their
irradiation tests they used a Co$^{60}$ source.

\section{Detector Production}
\begin{figure*}
\begin{center}
\epsfig{file=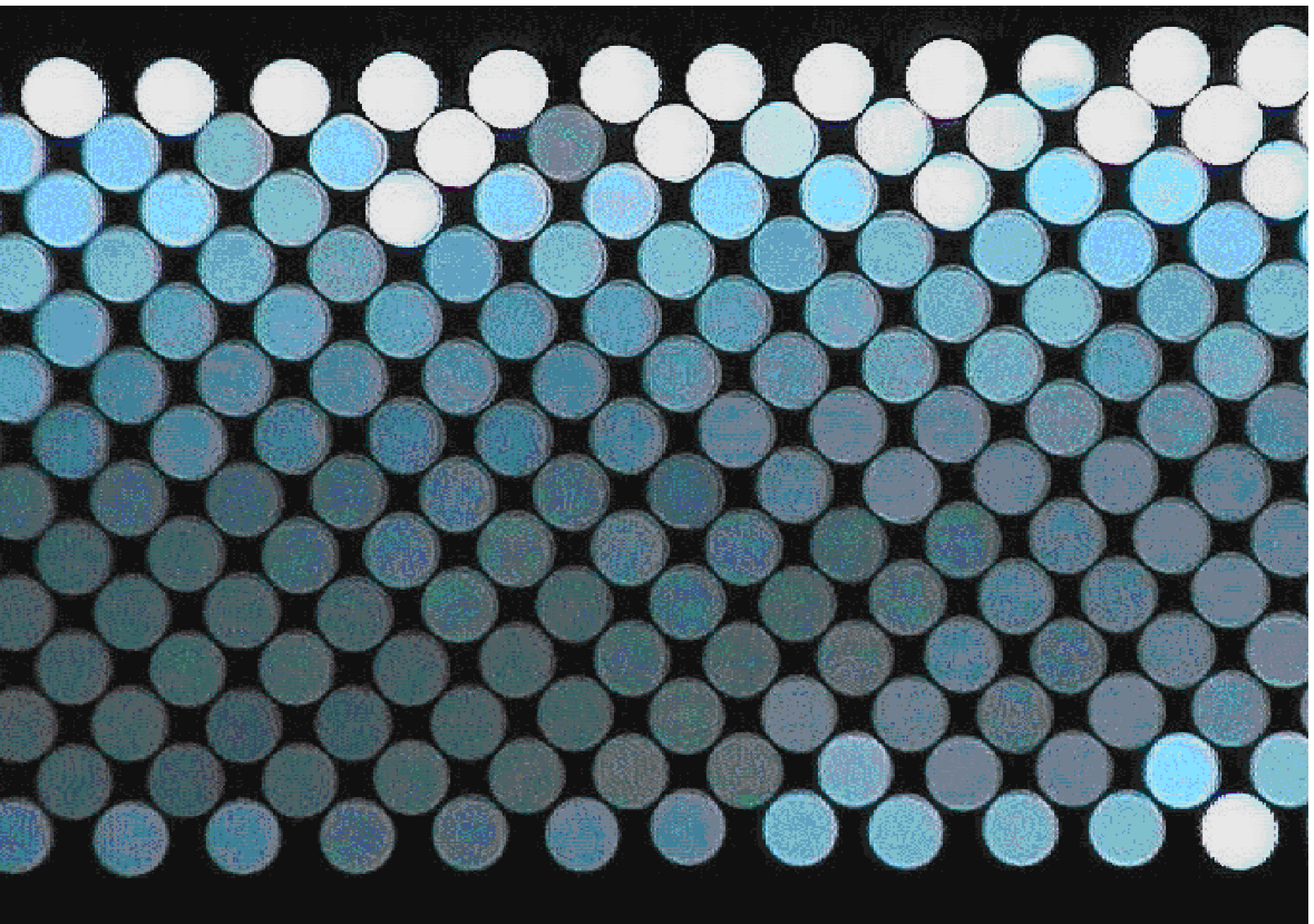,height=5cm}
\caption{\label{bilde} Photograph of part of the polished endface of 
the Zeuthen prototype detector}
\end{center}
\end{figure*}

Using winding technology as developed for the CHORUS experiment
\cite{lit10}
we built a detector production chain at our institute. A drum of 80 cm
diameter allows to produce five detectors at once. The production time
for winding one drum is about 14 h. Sticking the fibers to the
connector holes was still done by hand and rather time consuming. A
part
of the polished endface of one of our detectors is shown in fig. \ref{bilde}.

Two other detector prototypes are ordered from
industry. GMS--Berlin\footnote{GMS - Gesellschaft f\"ur Mess- und
 Systemtechnik mbH, Rudower Chaussee 5, 12489 Berlin, Germany}
followed a technology proposed by the university of Heidelberg
\cite{lit9}
mounting single layers on top of each other using epoxy glue. Each
layer is prepared on a v-grooved vacuum table. One layer per day can
be produced in this case. The connector is here also added by hand. 
The production technology used by KURARAY is unknown to us.

\begin{figure*}[t]
\begin{center}
\epsfig{file=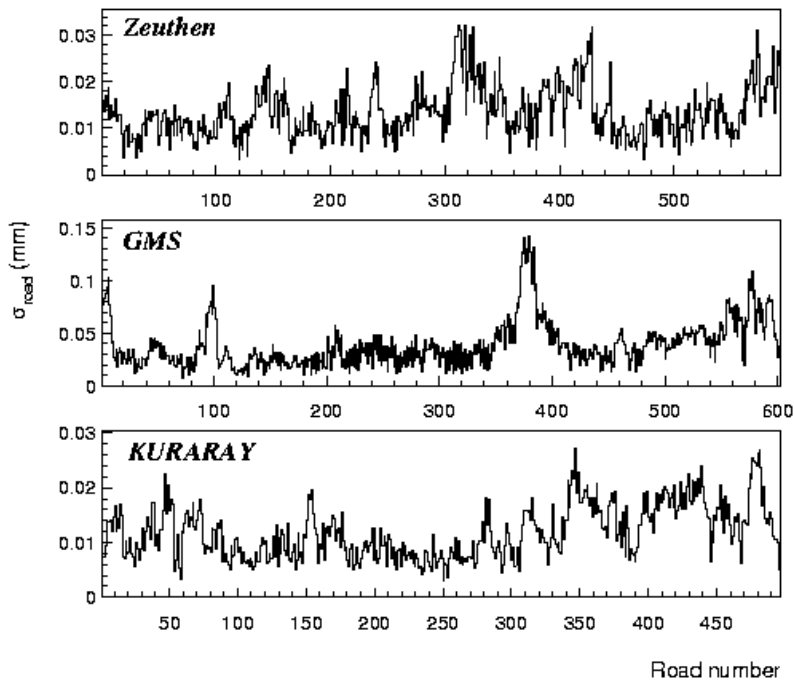,width=12cm}
\vspace*{-1cm}
\caption{\label{bildf} Deviation of fibers from ideal position per fiber road
            for the prototype detectors from Zeuthen, GMS and KURARAY}
\end{center}
\end{figure*}

To get the precision of the detector geometry quantified we
measured the coordinates of all fibers of the polished endface of the
three detectors. In fig. \ref{bildf} the deviation from the ideal position is
given per fiber road. Some stronger local effects are
visible. Averaging these results characteristic accuracies of
20 $\mu$m, 50 $\mu$m and 10 $\mu$m are calculated for the Zeuthen, GMS
and KURARAY detectors respectively.

\section{Opto--Electronic Readout}
The 640 channel scintillating fiber detectors are coupled via plastic
connectors to ten 3 m long light guide cables of 64 clear fibers  of
1.7 mm diameter. Each cable is coupled to a Hamamatsu 64 channel
position sensitive photomultiplier R5900 - M64. The coupling is done
using a plastic mask, fitting the anode pattern of the PSPM with $2
\cdot 2$ mm$^2$ pixel size. A detailed description of the
photomultiplier is given by the producer in \cite{lit11}.

The pulse rise time of the PSPM under study is a few
nanoseconds. Therefore the time behaviour of the fiber detector will
be mainly restricted by the readout electronics. For a high rate
experiment the signals have to be digitized and pipelined in an early
stage. We will not treat this problem here instead we try to measure
the total light amplitude as seen by the PSPM. For this purpose we
used the readout boards originally developed for the L3 fiber detector
\cite{lit12}. Like in previous beam tests \cite{lit13}, \cite{lit14} the serial output of these
boards is digitized by a SIROCCO--II flash--ADC using a VME OS9 online
data taking system with CAMAC interface.

We got the first prototypes of the PSPM just before the testrun
started. Finally we used only one exemplar for the beam
measurements. The corresponding light guide cable could be moved to
all ten connector places across the detectors. Mechanical precision
was guaranteed using precise pins and holes for each position. 
Using 20 nsec long light pulses of different intensity produced by a
LED coupled to a fiber we made an extensive check of the behaviour of
the PSPM in combination with our readout electronics. In the meantime
five other exemplars of the photomultiplier have been studied with
similar results.

\begin{figure*}[t]
\begin{minipage}[t]{6.5cm}
\epsfig{file=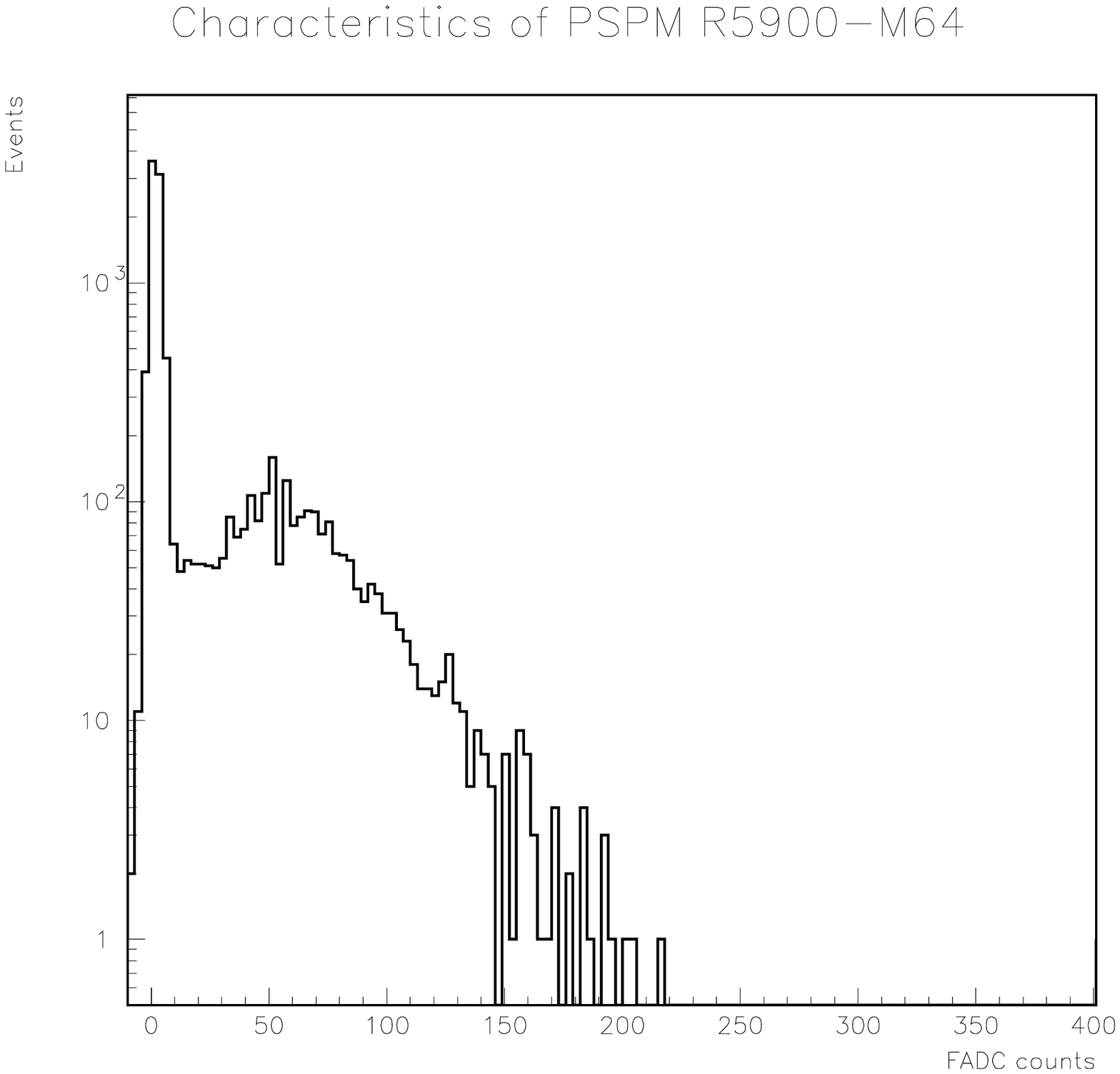,width=6.5cm}
\vspace*{-2cm}
\caption{\label{bildg} Flash ADC spectrum for one pixel of the 64 channel
            photomultiplier R5900-M64 illuminated with weak light
            signals from a LED}
\end{minipage}
\hfill
\begin{minipage}[t]{6.5cm}
\epsfig{file=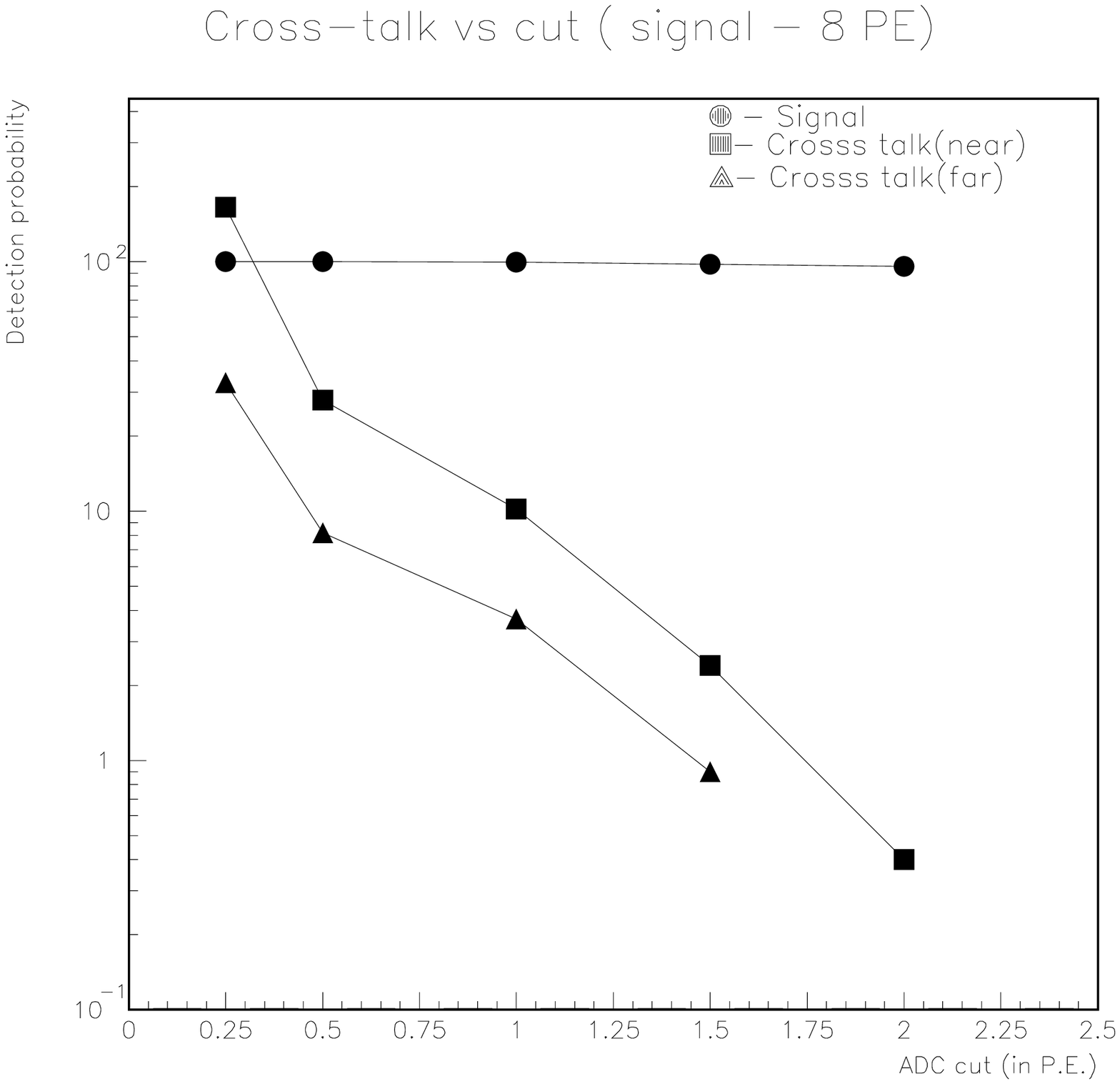,width=6.5cm}
\vspace*{-2cm}
\caption{\label{bildh} Detection probability for light signals corresponding to
            about 8 photoelectrons in dependence of a threshold cut
            given in numbers of photoelectrons for the illuminated
            pixel (circles), cross talk from direct neighbouring
            pixels (squares) and cross talk from distant ones
            (triangles)}
\end{minipage}
\end{figure*}

In fig. \ref{bildg} we show the amplitude distribution for one pixel
illuminated with weak light signals. Beside a narrow pedestal a clear
one photoelectron peak is seen. The gain of the PSPM comes out to be
independent of the number of incoming photons up to more than 10
photoelectrons and linear in the high voltage range between 800 and
980 V. This allows to calibrate the light output measured in 
FADC-channels to the corresponding average number of photoelectrons. What
has to be taken into account in addition in this case is the different
sensitivity of the PSPM pixels. We measured it, putting the same
amount of light to each channel and found a variation by about a
factor of two.

The applicability of a PSPM for fiber detector readout is limited by
its cross talk properties, i.e. the response of other pixels than the
illuminated one to the incoming light signal. In case of the
R5900-M64, Hamamatsu has 
decreased the cross talk already considerably by decreasing the
thickness of the entrance window from 1.3 mm to 0.8 mm. Nevertheless the
cross talk depends strongly on the threshold cut applied and is still
about 10 \% for a one photoelectron cut and typical light signals as can be
seen from fig. \ref{bildh}. From the figure one observes also that most of the
cross talk comes from the nearest neighbours of the directly
illuminated pixel. Using an appropriate hit selection procedure this
cross talk can be strongly suppressed. If for instance a local maximum
selection is used it vanishes completely \cite{lit13}, \cite{lit14}.

\begin{figure*}[b] 
\begin{minipage}[b]{6.5cm}
\hspace*{0.3cm}
\epsfig{file=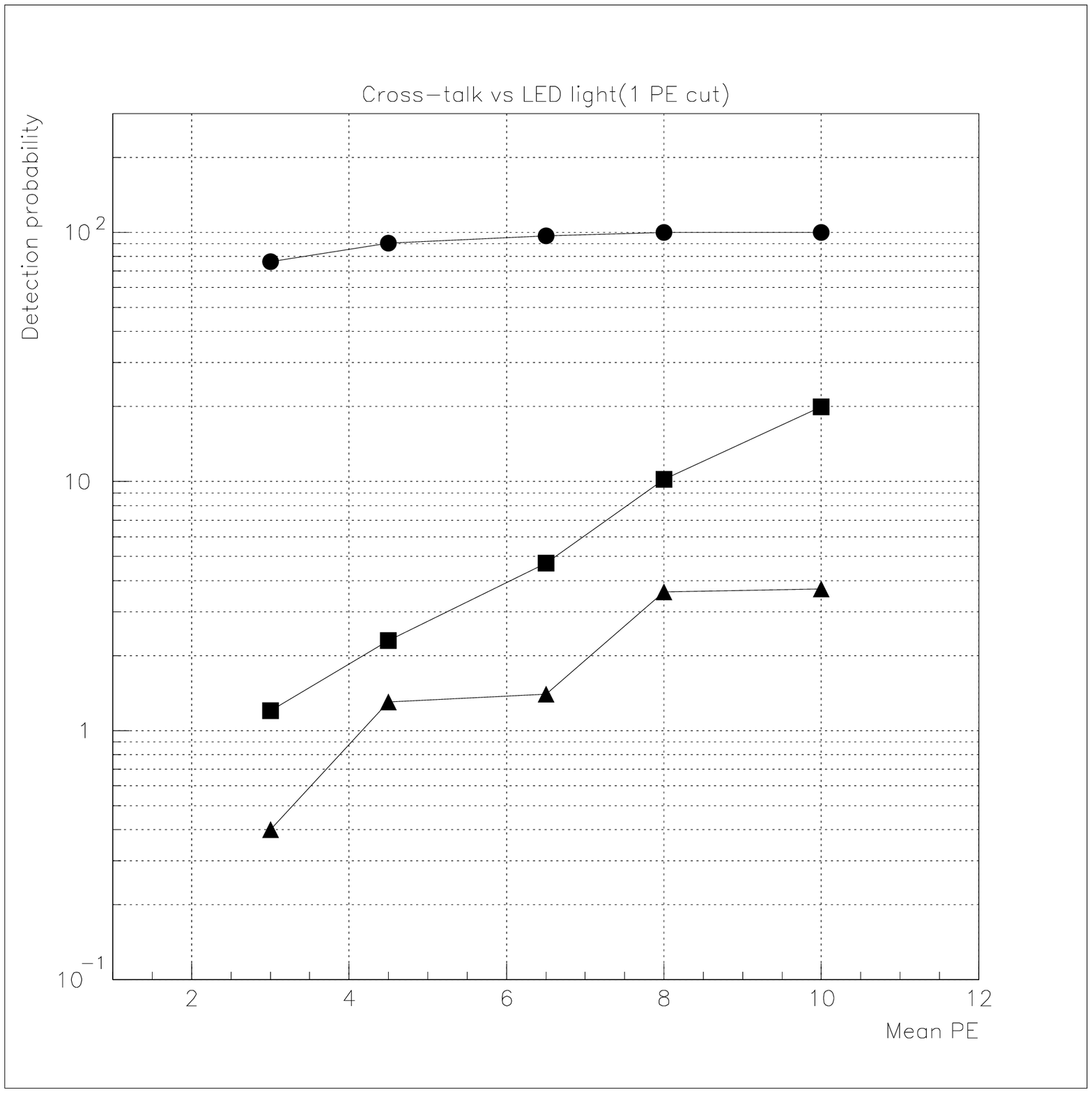,width=6cm}
\vspace*{-1cm}
\caption{\label{bildi} Detection probability for signal (circles),
  near (squares) and far (triangles) cross talk for incoming light of
           different intensity given in number of photoelectrons.
            An one photoelectron cut has been applied to the data.}
\end{minipage}
\hfill
\begin{minipage}[b]{6.5cm}
\hspace*{0.2cm}
\epsfig{file=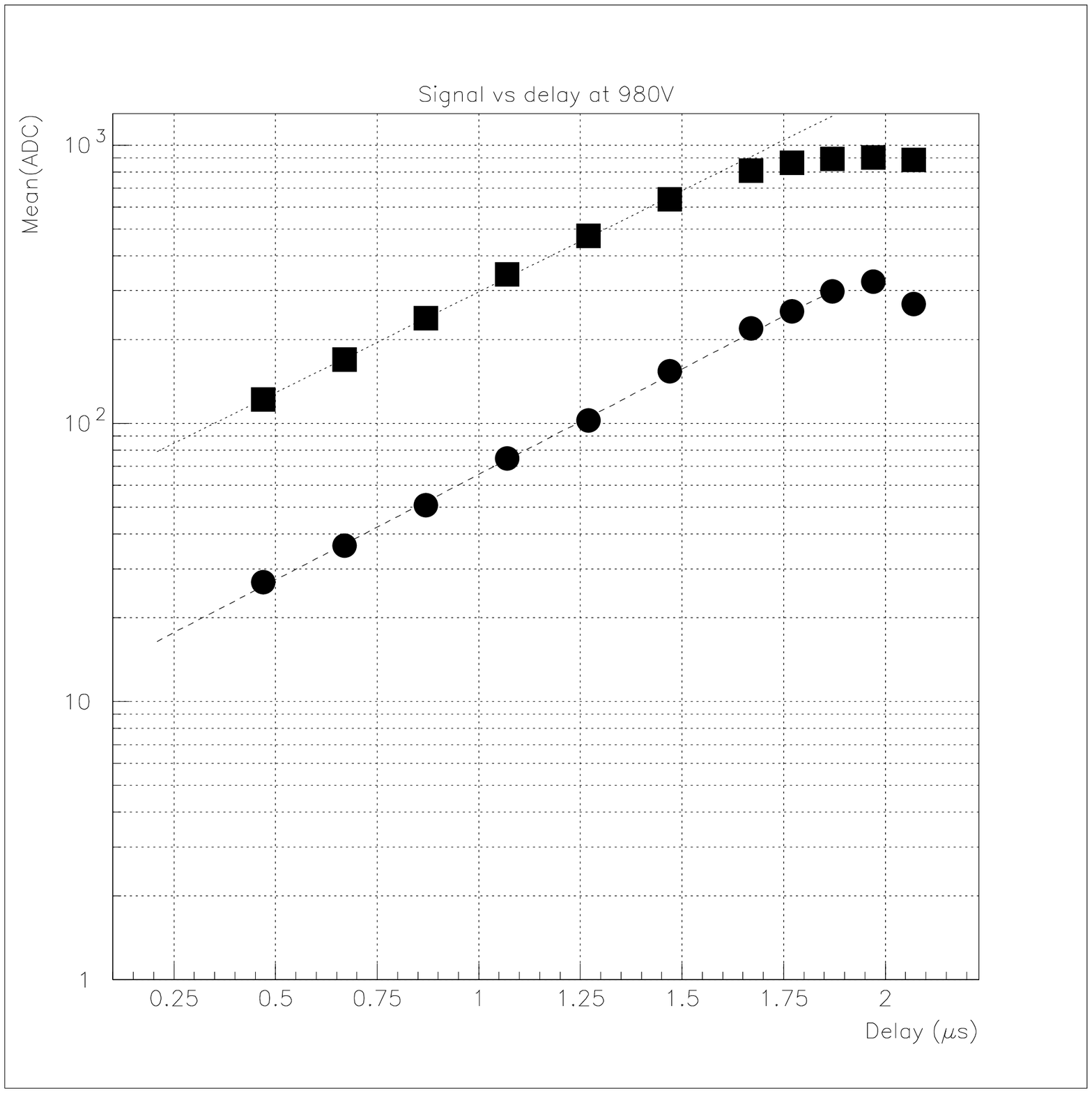,width=6cm}
\vspace*{-1cm}
\caption{\label{bildj} Average number of FADC - channels in dependence of the
            arrival time of the corresponding trigger in a 2 $\mu$sec
            gate for constant  light signals of two different
            intensities (squares and circles)}
\end{minipage}
\end{figure*}

Whereas the amount of cross talk depends only weakly on the PSPM
high voltage, it is strongly correlated to the amount of incoming light
to a pixel (see fig. \ref{bildi}). The diameter of the light guide fibers
connected to the pixels starts to influence the cross talk behaviour
above 1.5 mm.

Our fiber detector data taking in the testbeam has been 
synchronized with respect to data taking with a four plane Microstrip
Detector Hodoscope (MSD). To increase the final data rate for the
two systems with free running clocks a long trigger gate
of 2 $\mu$sec was chosen. As found later, that leads already to discharge
effects for early PSPM signals. In fig. \ref{bildj} the flash ADC response to a
constant light signal arriving at different trigger times is shown. For
our testrun data the arrival time of a trigger is unknown but
uniform. Correspondingly we had to unfold all flash ADC spectra 
using the behaviour measured in fig. \ref{bildj} to determine the correct
average number of photoelectrons.

\vspace*{-1cm}

\vspace*{-1cm}

\clearpage

\newpage

\section{Testrun Setup}
\begin{figure*}[h] 
\begin{center}
\vspace*{0.5cm}
\epsfig{file=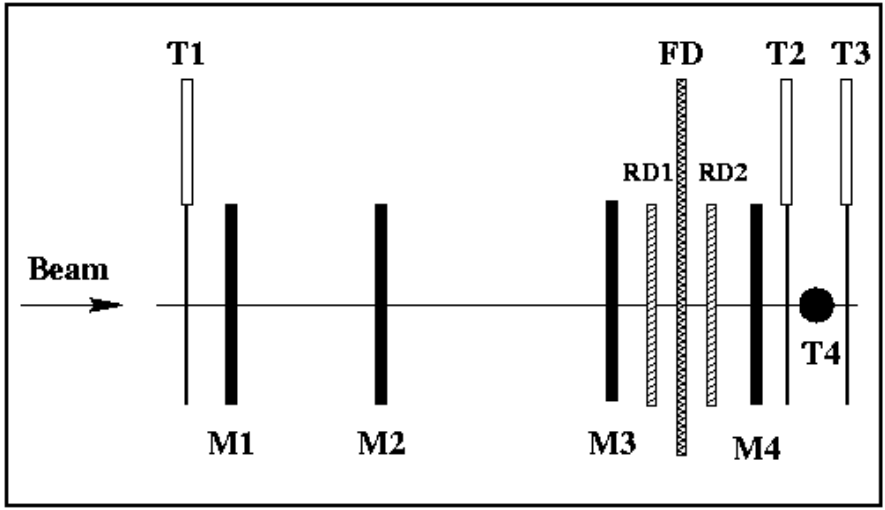,height=7cm}
\caption{\label{bildk} Schematic view of the testrun setup. M1-M4: silicon
            microstrip detectors, RD1 and RD2: fiber reference
            detectors, T1-T4: trigger paddles, FD: fiber detector
            to be tested}
\end{center}
\end{figure*}

  In a two weeks testrun in October 1997 the three available detector
prototypes have been studied in the 3 GeV electron beamline T21 at
DESY. More than 5*10$^5$ triggers have been recorded  under several conditions in various
runs. The setup is schematically drawn in fig. \ref{bildk}.

A trigger system of four scintillator paddles T1 - T4 selects by
4-fold coincidence a beam region of 2*2 cm$^2$. The corresponding
trigger rate was 50 kHz, leading to a final data acquisition rate of
about 5 Hz.

Two fiber reference detectors RD1 and RD2 fixed to the same support
as the MSD hodoscope paddles are used mainly to perform an
independent alignment of these components. Each fiber reference
detector is made out of two double layers of 32 fibers of 1 mm
diameter. The total intrinsic resolution of the system is 72 $\mu$m.

The MSD hodoscope consists of four paddles with 320 or 640 strips of
100 $\mu$m pitch. The intrinsic resolution for one hit clusters is about
30 $\mu$m.
 
The distribution of differences of coordinates measured by the fiber
reference detector and the corresponding MSD paddle has been used to
align the measuring system. Alignment errors of 58 $\mu$m, 70 $\mu$m,
13 $\mu$m, and 10 $\mu$m have been found for the paddles 1 to 4. The result reflects
problems of the geometrical arrangement with respect to the beam
divergence. The MSD prediction precision at the fiber detector
position has been calculated from fits of four hit tracks to be 75
$\mu$m.

The fiber detector FD is movable in x and y direction perpendicular
to the beam. About 30 different positions have been used for
measurements. The alignment of the fiber detector with respect to the
MSD-coordinate system has been done separately for every position with
a precision of about 10 $\mu$m.

\section{Results from Testrun Data}
All testrun data have been sensitivity corrected corresponding to
the correction table measured before for the PSPM operating with a high
voltage of \linebreak 980 V. A noise threshold of 35 FADC channels has been
applied afterwards leaving 1 $^o/_{oo}$ efficiency for  a noise hit per beam
trigger. The threshold to select events with more than one
photoelectron seen is chosen to be 150 FADC channels at this high
voltage with the one photoelectron peak observed at 100 FADC channels.

\begin{figure*}[b]
\begin{center}
  \epsfig{file=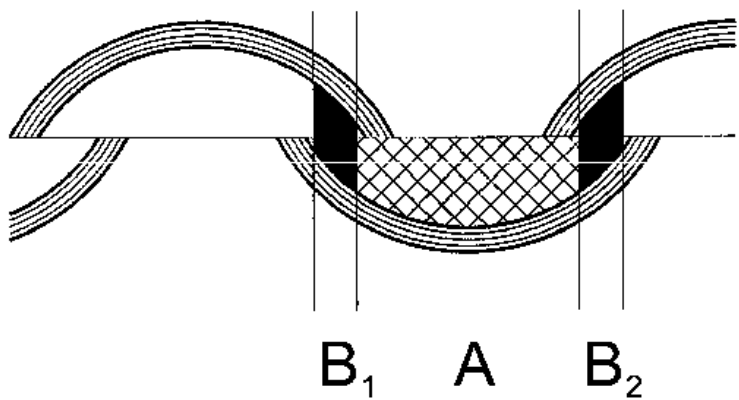,width=11cm}
\vspace*{-5cm}
\caption{\label{bildl} Illustration of the different zones A and B for
            overlapping fiber roads leading to one or two hit events}
\end{center}
\end{figure*}

For the selection of hitted fiber roads in principle a simple cut
excluding single photoelectron hits would be sufficient to reduce
background from cross talk and other sources. Results can be improved,
if a local maximum search is possible removing all nearby cross talk
hits \cite{lit13}, \cite{lit14}. In the following we will use both
selection criteria in common.

\begin{figure*}[t]
\begin{center}
\epsfig{file=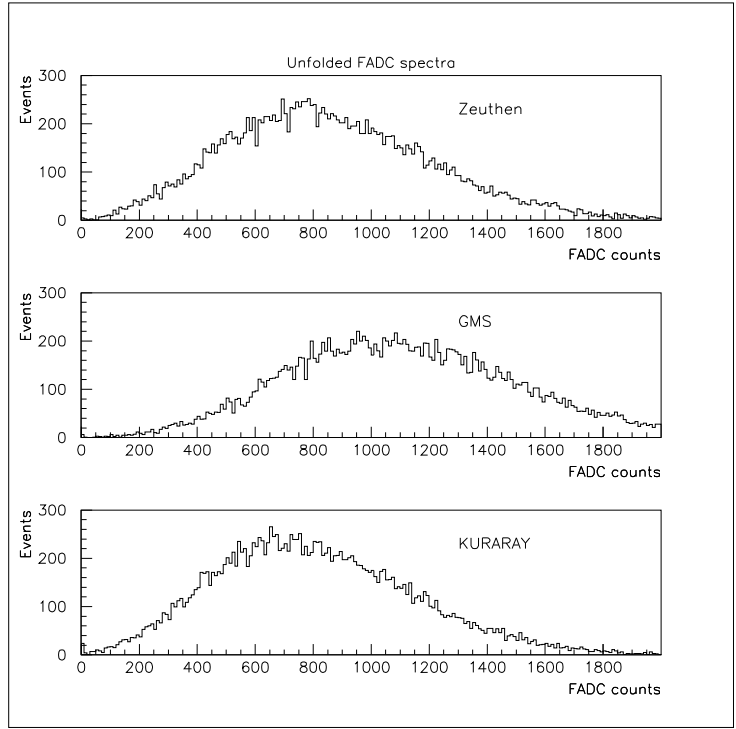,height=12cm}
\vspace*{-1cm}
\caption{\label{bildm} Unfolded light output spectra as measured by the FADC
            for the detectors from a.) Zeuthen, b.) GMS and c.)
            KURARAY}
\end{center}
\end{figure*}

After selecting hits, a cluster algorithm combines neighbouring ones
to a track cluster. As can be seen from fig. \ref{bildl} 
from geometry one or two fiber roads could be
hitted by a particle crossing perpendicular to the detector plane . 

\begin{table*}[b]
\caption{\label{tab1} Average number of photoelectrons $ <N_{pe}>$, efficiency
  $\epsilon$ and resolution  $\sigma$ for the fiber detectors produced
  by Zeuthen, GMS and KURARAY}
\vspace*{0.5cm}
\begin{center}
\begin{tabular}{|c|c|c|c|} \hline
detector & $<N_{pe}>$ &  $\epsilon/\%$ & $\sigma/\mu$m \\ \hline
\hline
Zeuthen  &  8.2      &     99.4  & 94           \\ \hline
GMS      & 10.1      &     99.9  &   90         \\ \hline
KURARAY  &  7.5      &     99.0  &     88       \\  \hline
\end{tabular}
\end{center}
\end{table*}

In figures \ref{bildm} a-c the unfolded FADC spectra are shown for
the detectors produced by Zeuthen, GMS and KURARAY respectively. Using
the described 
calibration leads to an average number of photoelectrons between eight
and ten and correspondingly efficiencies above 99 \% if
the one photoelectron cut is applied (see table \ref{tab1}). The larger values
for the GMS detector seem to be due to the very good quality of the 
polished and mirrored detector endface.

\begin{figure*}[t]
\begin{center}
\epsfig{file=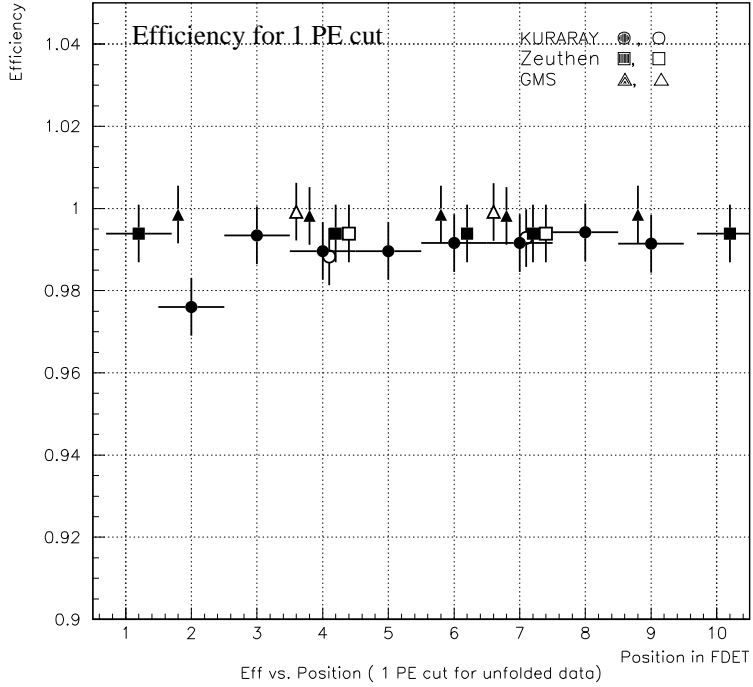,width=10cm}
\vspace*{-1.8cm}
\caption{\label{bildn} Efficiency at different positions of the fiber
            detector prototypes from Zeuthen (squares), GMS
            (triangles) and KURARAY (circles) averaged for 64 fiber
            roads. Particles hit 5 cm from the near (open symbols) or
            far (filled symbols) end of the ordered detector part. An
            one photoelectron cut has been applied to all data.}
\end{center}
\end{figure*}

As can be seen from fig. \ref{bildn} the efficiency varies only weakly for
different positions of the light guide cable, i.e. different fiber
roads across the detector. The variation of the efficiency along the
fibers is also marginal. However the distance between near and far
measurements is only about 15 cm.

\begin{figure*}
\begin{center}
\epsfig{file=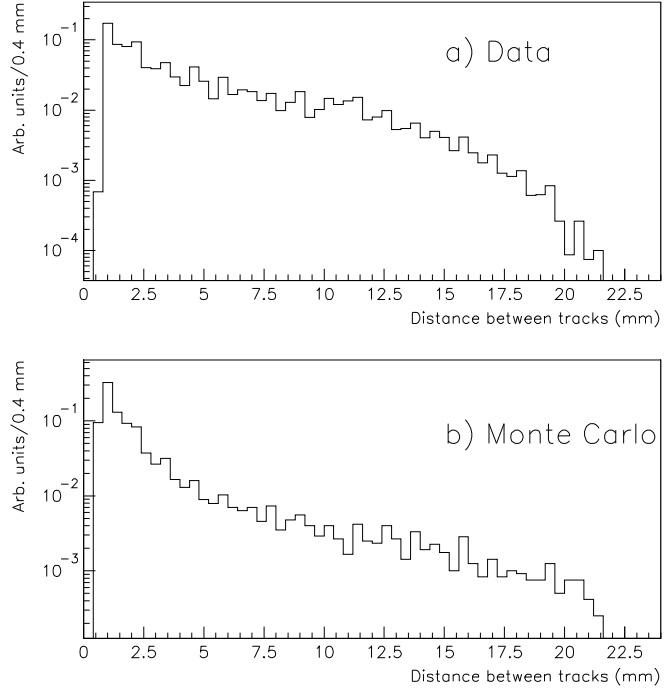,width=10cm}
\vspace*{-1.8cm}
\caption{\label{bildo} Distribution of distances between tracks in multitrack
  events after an one photoelectron cut on the light output has
            been made for a.) data and b.) Monte Carlo simulation.}
\end{center}
\end{figure*}

Expecting in principle only one incoming particle from beam triggers
one observes in the data after the one photoelectron cut still 10 \% of
events with more than one track cluster in the fiber detector.  A
rough GEANT Monte Carlo simulation including only the detector
material comes up with about \linebreak 4 \% additional hits produced by matter effects.
These hits increase the number of two hit track clusters and lead to
additional "matter" tracks. 2 \% of such tracks are observed.
 Fig. \ref{bildo} shows the distance
between tracks in multi-track events for data and Monte
Carlo. Comparing both one can not exclude cross talk at the percent
level. Possible sources are accidentals (noise), the PSPM glass
window, fibers in the detector, connectors or light guide cables.

To calculate the intrinsic resolution of our fiber detector geometry
one has to keep in mind again that the fiber roads have two regions A and B
where one or two of them should give light signals for throughgoing
particles (see fig. \ref{bildl}). With a fiber diameter of 
480 $\mu$m, a cladding thickness of 30 $\mu$m  and a pitch of 340
$\mu$m 
values of 260 $\mu$m and  80 $\mu$m are derived for the widths of the
regions A and B. Keeping in mind that the one photoelectron cut
excludes low light signals one calculates the values given in table
\ref{tab2} for the width of the regions, the fraction of crossing particles
and the corresponding resolution. Combining these numbers an overall
intrinsic resolution of 69 $\mu$m could be expected.

\begin{table*}
\caption{\label{tab2} Width, fraction of appearance, intrinsic geometrical
  resolution and resolution measured in beam test for the two regions 
of a fiber road where one or two hits are produced by a crossing particle.}
\vspace*{0.5cm}
\begin{center}
\begin{tabular}{|c|c|c|c|c|} \hline
region  & width/$\mu$m & fraction/$\%$ & $\sigma_{geom}$/$\mu$m &
$\sigma_
{FD}$ /$\mu$m \\ \hline \hline
A    & 280 & 82 & 81 & 97 \\ \hline
B    &  60 & 18 & 17 & 30 \\ \hline
A + B  & 340 & 100& 69 & 86 \\ \hline
\end{tabular}
\end{center}
\end{table*}

\begin{figure*}
\begin{center}
\epsfig{file=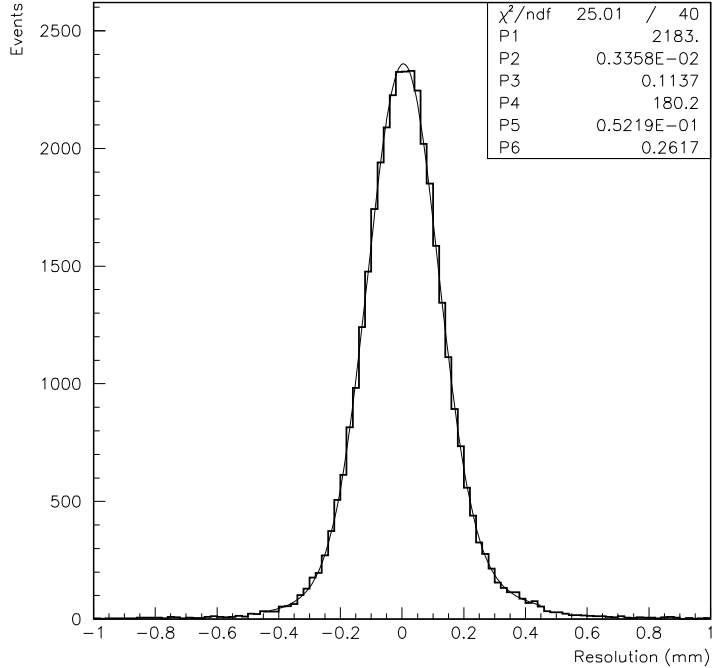,width=10cm}
\vspace*{-1.8cm}
\caption{\label{bildp} Distribution of the difference of coordinates
  measured by the fiber detectors and the MSD hodoscope after an one
            photoelectron cut has been applied. The curve is the
            result of a fit with two gaussian functions.}
\end{center}
\end{figure*}

The difference between the coordinates of a crossing particle
measured with the fiber detector and with the MSD hodoscope is
distributed in fig. \ref{bildp} for all data of all runs. A fit with two gaussian functions yields a
resolution of $\sigma = 114 \mu$m for the main component. Unfolding the MSD
precision of 75 $\mu$m a fiber detector resolution of 86 $\mu$m is calculated,
near to the theoretical expectation.

Separating events with one and two fiber roads hit allows to
calculate the resolution for tracks crossing regions A and B. The
values given in the last column of table \ref{tab2} are also close to the
theoretical estimates. Only the double road hit distribution needs
two gaussian functions to be described reasonably. The second
component with a large width is mainly due to matter effects like
delta-electrons which lead to additional hits in neighbouring fibers.

\begin{figure*}
\begin{center}
\epsfig{file=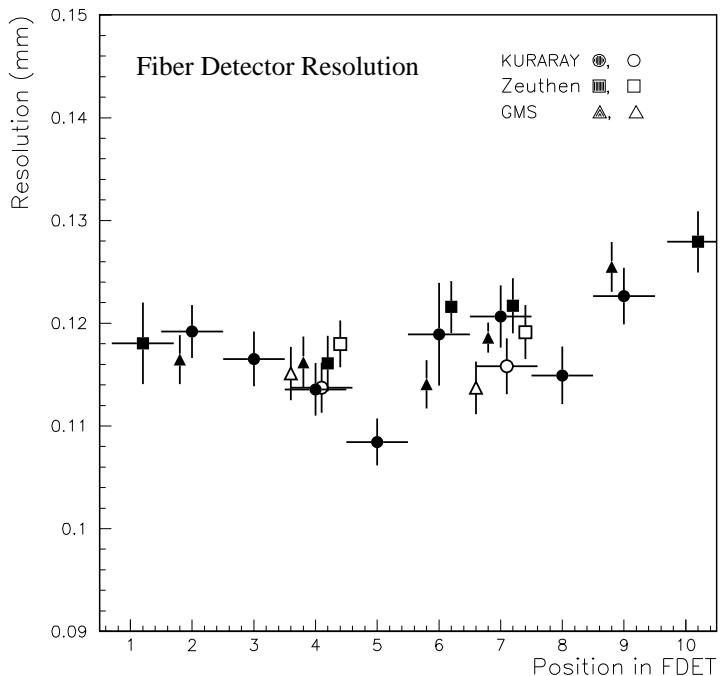,width=10cm}
\vspace*{-1.8cm}
\caption{\label{bildq} Resolution at different positions of the fiber
            detector prototypes from Zeuthen (squares), GMS
            (triangles) and KURARAY (circles) averaged for 64 fiber
            roads. Particles hit 5cm from the near (open symbols) or
            far (filled symbols) end of the ordered detector part. An
            one photoelectron cut has been applied to all data.}
\end{center}
\end{figure*}

The resolution has also been measured for different positions in
different detectors (see fig. \ref{bildq} and table \ref{tab1}). No remarkable variations
were observed. The statistics and precision of measurements are not
sufficient to establish local effects due to problems in mechanical
detector production.

A few additional phenomena which may influence the tracking
precision in an experiment have also been considered during the
testrun. All prototypes have "unordered" fiber regions of about 7cm
near to the 640 channel connector. Throughgoing particles are
registrated in this region with a precision of about 1 mm.

The 640 clear fibers of 1.7 mm diameter and 3 m length give a good
possibility to produce Cerenkov light if crossed by charged
particles. That depends strongly on the light guide arrangement: the
largest effect can be expected near to the connector. Exposing
this region to the electron beam, less than 1 \% of crossing particles
produced hits after applying the one photoelectron cut.

Particle showers from nearby material increase background and
occupancy and decrease the detector resolution. A small effect of this
kind may be seen already in fig. \ref{bildq} comparing the values measured for
the detector boundaries with those in the central region.

\section{Summary}
Three fiber detector prototypes have been tested. They are made out
of 640 overlapping roads of seven 480 $\mu$m fibers coupled to 1.7 mm
diameter light guides of 3 m length readout with 64 channel
photomultipliers.

For all three detectors an efficiency of about 99 \% and a resolution
better than 100 $\mu$m have been measured in an exposure to a 3 GeV
electron beam at DESY.

These results together with radiation hardness studies of the used
fiber materials seem to make it possible to use a corresponding
detector in a high rate experiment like HERA-B.

The readout of such a detector - connectors, light guides, mask
arrangement, front end electronics - has to be adapted with care to
the specific demands of the application.

\section*{Acknowledgement}
Part of this work was done in close collaboration with groups from the
universities of Heidelberg and Siegen. We want to thank our colleagues
for their good cooperation and many fruitful discussions.

In particular we thank Prof. F.Eisele for providing the KURARAY 
detector, bought by Heidelberg University, for the beam tests.

The fiber irradiation tests were possible only due to the kind support
of the Hahn-Meitner-Institute Berlin. We are deeply indebted to the
ISL accelerator team and want to thank in particular
Dr. D. Fink, Dr. K. Maier and Dr. M. M\"uller from HMI and Prof. H.A. Klose
from GMS for a lot of practical help.

We acknowledge the benefit from the DESY II accelerator crew and
the test area maintainance group.


\begin{thebibliography}{99}
\bibitem{lit1}Ansorge, R., et al., {\it NIM} {\bf 265}, 33 (1988)
\bibitem{lit2}Annies, P., et al., {\it NIM} {\bf A367}, 367 (1995)
\bibitem{lit3}Bross, A.D., {\it Nucl. Phys. B (Proc.Suppl.)} {\bf 44},
        12 (1995)\\
       Adams,D.,et al., {\it Nucl. Phys. B (Proc.Suppl.)} {\bf 44},
       332 (1995)
\bibitem{lit4}B\"ahr, J., et a., {\it Proceedings of the 28th Intern.
              Conf. on High Energy Physics, Warsaw, Poland, 1996,
              eds. Z.Ajduk,A.K.Wroblewski} {\bf V. II}, p. 1759
\bibitem{lit5}Lohse, T., et al., {\it HERA-B Technical Proposal,
              DESY-PRC} {\bf 94/02} (1994)
\bibitem{lit6}Ferro--Luzzi, M., et al., contribution presented by
A.Gorin, to appear in Proceedings of the
workshop SCIFI97, Notre Dame, USA, 1997

\bibitem{lit7}Yoshizawa, Y., et al., to appear in Proceedings of the
workshop SCIFI97, Notre Dame, USA, 1997
\bibitem{lit8}Aschenauer, E.C., et al., preprint {\it DESY} {\bf
  97-174} (1997)
\bibitem{lit9}Dreis, B., et al., preprint {\it DESY} {\bf 98-049} (1998)
\bibitem{lit10}Nakano, T., et al., {\it Proceedings of the workshop
  SCIFI93, Notre Dame, USA, 1993, eds. A. Bross, R. Ruchti, M. Wayne},
p. 525
\bibitem{lit11} Yoshizawa, Y., et al., Hamamatsu technical
information, No. TPMH 9002E01
\bibitem{lit12} Akbari, H., et al., {\it NIM} {\bf A302}, 415 (1991)
\bibitem{lit13} B\"ahr, J., et al.,  Proceedings of the workshop
          SCIFI93, Notre Dame, USA, 1993, eds. A. Bross, R. Ruchti,
          M. Wayne, p.578
\bibitem{lit14} B\"ahr, J., et al., {\it NIM} {\bf A371}, 380 (1996)
\end{thebibliography}
\end{document}